%% file: main.tex
\documentclass[10pt,onecolumn,romanappendices]{IEEEtran} 
\usepackage{setspace}
\usepackage{multirow}

\usepackage[left=1in,right=1in,top=1in,bottom=1in]{geometry}
\input{resources/latex_headers}

\title{\huge Multi-way Graph Signal Processing on Tensors: \\
\Large Integrative analysis of irregular geometries}

\author{Jay S. Stanley III, Eric C. Chi, and  Gal Mishne
\thanks{J. S. Stanley is with the Dept. of Mathematics, Yale University, New Haven, (e-mail: jay.stanley@yale.edu). E. C. Chi is with the Dept. of Statistics, NC State University, Raleigh, NC, (e-mail: eric\_chi@ncsu.edu). G. Mishne is with the Hal\i c\i o\u glu Data Science Institute, UC San Diego, La Jolla, CA (email: gmishne@ucsd.edu).}}

\date{}
\begin{document}

\maketitle

\vspace{-2cm}
\begin{abstract}
Graph signal processing (GSP) is an important methodology for studying data residing on irregular structures. 
As acquired data is increasingly taking the form of multi-way tensors, new signal processing tools are needed to maximally utilize the multi-way structure within the data. In this paper, we review modern signal processing frameworks generalizing GSP to multi-way data, starting from graph signals coupled to familiar regular axes such as time in sensor networks, and then extending to general graphs across all tensor modes. This widely applicable paradigm motivates reformulating and improving upon classical problems and approaches to creatively address the challenges in tensor-based data.  
We synthesize common themes arising from current efforts to combine GSP with tensor analysis and  highlight future directions in extending GSP to the multi-way paradigm.

\end{abstract}
\vspace{-0.3cm}
\section{Introduction}
\input{intro}
\vspace{-0.3cm}
\section{Single-way GSP}
\label{sec:gsp}
\input{GSP}
\vspace{-0.3cm}
\section{Extending GSP to multi-way spaces}
\label{sec:mwgsp}
\input{gsp_tensors2}
\input{extensions}
\vspace{-0.3cm}
\section{Signal processing on multi-way graphs}
\label{sec:mwsp}
\input{graph_regularize}
\vspace{-0.3cm}
\section{Future Outlook}
\label{sec:discussion}

As multi-way signal processing frameworks continue to mature, several challenges remain ahead. While novel techniques are continually introduced into single-way graph signal processing, one approach to developing multi-way techniques is to identify, extend, and adapt techniques which are particularly useful for multi-way signals. For instance, multi-way analysis on directed graphs will greatly broaden the versatility of MWGSP.  From a computational perspective, it is clear that the efficiency gains offered by the march of single-way GSP march towards fast transforms~\cite{frerix2019approximating} are compounded in the multi-way setting. 

From a theoretical perspective, open questions include
\begin{enumerate*} 
\item What additional advantages can be gained by treating classical domains as lying on graphs? 
\item How do we learn mode-specific or coupled graphs from data, in general and in dynamical settings? \item Are such tensor datasets typically low-rank or high-rank? \item How do we process data whose generative model is nonlinear across the different modes? 
\end{enumerate*}

From a practical perspective,
ongoing growth in computational power and  parallel computing 
have enabled large-scale analyses. The MWGSP framework can leverage these recent advances in computational building blocks. Nonetheless, there are existing computational challenges, such as applications requiring online real-time processing. 
Thus, future directions include developing online and distributed versions of multi-way graph signal processing, especially in the presence of large-scale data, where streaming solutions are necessary (the data does not fit in memory).
In addition, there is need for new optimization techniques to efficiently solve problems that combine tensors with graph-based penalties.
Deep learning is also emerging as a framework to learn rather than design wavelet-type filterbanks in signal processing 
and these approaches can be extended to the graph and multi-way settings to learn joint multiscale decompositions. 
Finally, as the GSP community continues to address real-world data domains such as climate, traffic, and biomedical research, inter-disciplinary collaboration is essential to define relevant problems and demonstrate significant utility of these approaches within a domain. 


\begingroup
\footnotesize{
\setstretch{1}
\bibliographystyle{IEEEtran} 
\bibliography{references.bib}
}
\endgroup

\end{document}

%% file: resources/latex_headers.tex
\usepackage[utf8]{inputenc}
\usepackage{amsmath,amssymb,amsfonts,mathtools,scalerel}
\usepackage{mleftright}
\usepackage[mathscr]{eucal} 
\usepackage{bm} 
\usepackage{soul}
\usepackage{color}
\usepackage{comment}
\usepackage[inline]{enumitem}
\usepackage{multirow}
\usepackage{mdframed}

\usepackage{graphicx}

\usepackage[numbers,square,compress]{natbib}
\usepackage{hyperref}
\usepackage[capitalise,nameinlink]{cleveref} 
\usepackage{algpseudocode}
\usepackage{algorithm}

\makeatletter
\let\MYcaption\@makecaption
\makeatother
 
\usepackage[style=base, font=footnotesize,labelformat=simple]{subcaption}

\makeatletter
\let\@makecaption\MYcaption
\makeatother

\crefname{algorithm}{Alg.}{Algs.}
\Crefname{algorithm}{Algorithm}{Algorithms}
\Crefname{chapter}{Chapter}{Chapters}
\crefname{chapter}{Chap.}{Chaps.}
\Crefname{section}{Section}{Sections}
\crefname{section}{Sec.}{Secs.}
\Crefname{figure}{Figure}{Figures}
\crefname{figure}{Fig.}{Figs.}
\Crefname{table}{Table}{Tables}
\crefname{table}{Table}{Tables}
\makeatletter

\def\refstepcounter@noarg#1{%
  \cref@old@refstepcounter{#1}%
  \cref@constructprefix{#1}{\cref@result}%
  \@ifundefined{cref@#1@alias}%
    {\def\@tempa{#1}}%
    {\def\@tempa{\csname cref@#1@alias\endcsname}}%
  \protected@edef\cref@currentlabel{%
    [\@tempa][\arabic{#1}][\cref@result]%
    \noexpand\@currentlabel}} 

\def\refstepcounter@optarg[#1]#2{%
  \cref@old@refstepcounter{#2}%
  \cref@constructprefix{#2}{\cref@result}%
  \@ifundefined{cref@#1@alias}%
    {\def\@tempa{#1}}%
    {\def\@tempa{\csname cref@#1@alias\endcsname}}%
  \protected@edef\cref@currentlabel{%
    [\@tempa][\arabic{#2}][\cref@result]%
    \noexpand\@currentlabel}}

\makeatother    

\def\TODO#1{{\color{red} \textbf{#1}}} 

\definecolor{blue}{rgb}{0.2,0.5,0.7}
\definecolor{green}{rgb}{0.3,0.68,0.29}
\definecolor{purple}{rgb}{0.6,0.31,0.64}

\newcommand{\Tra}{^{\sf T}} 
\def\vec{\mathop{\rm vec}\nolimits}

\newcommand{\V}[1]{{\bm{\mathbf{\MakeLowercase{#1}}}}} 
\newcommand{\Vn}[2]{\V{#1}^{(#2)}} 

\newcommand{\M}[1]{{\bm{\mathbf{\MakeUppercase{#1}}}}} 

\newcommand{\Mn}[2]{\M{#1}^{(#2)}} 

\newcommand{\T}[1]{\boldsymbol{\mathscr{\MakeUppercase{#1}}}} 

\newcommand{\Kron}{\otimes} 

\newcommand{\Real}{\mathbb{R}}
\newcommand{\Complex}{\mathbb{C}}

\newcommand{\G}{\mathcal{G}}
\newcommand{\Cart}{\times} 
\def\Bsquare{\mathop{\mathord{\scalerel*{\Box}{\sum}}}}  
\newcommand{\GCart}{\Bsquare} 
\def\Msquare{\mathop{\mathord{\scalerel*{\Box}{+}}}}  
\newcommand{\Gcart}{\Msquare} 
\newcommand{\Ec}{\mathcal{E}}
\newcommand{\Vc}{\mathcal{V}}
\def\Bboxtimes{\mathop{\mathord{\scalerel*{\times}{\sum}}}}


%% file: intro.tex
Over the past decade, graph signal processing (GSP)~\cite{ortega2018graph} has laid the foundation for generalizing classical Fourier theory as defined on a regular grid, such as time, to handle signals on irregular structures, such as networks. GSP, however, is currently limited to single-way analysis:  
graph signals are processed independently of one another, thus ignoring the geometry between multiple graph signals. 
In the coming decade, generalizing GSP to handle multi-way data, represented by multidimensional arrays or \emph{tensors}, with graphs underlying each axis of the data will be essential for modern signal processing. 
This survey discusses the burgeoning family of \emph{multi-way graph signal processing} (MWGSP) methods for analyzing data tensors as a dependent collection of axes. 


To introduce the concept of \emph{way}, consider a network of $N$ sensors each measuring a signal sampled at $T$ time points. On the one hand, classic signal processing treats these signals as a collection of $N$ independent 1D time-series 
ignoring the relation structure of the graph. On the other hand, the standard GSP perspective treats the data as a collection of $T$ independent 1D graph signals 
that describe the state of all sensors for a given time point $t_j \in T$.
Both are single-way perspectives that ignore the underlying geometry of the other way (also referred to as \emph{mode}).
The recent time-vertex (T-V) framework~\cite{grassi2017time,romero2017kernel} unifies these perspectives to form a dual-way framework that processes graph signals that are time-varying\footnote{Note that the graph itself is static while the signals are time-varying}, thus bridging the gap between classical signal processing and GSP.
While one of the axes of a T-V signal is a regular grid, time, in general a regular geometry may not underlie any of the ways of the data, e.g.\@ genes and cells in sequencing data or users and items in recommendation systems~\cite{rao2015collaborative,Chi2017a,Mishne2017}. 
Thus, the T-V framework is a subset of a more general MWGSP framework that considers the coupling of multiple  geometries, whether predefined temporal or spatial axes, or irregular graph-based axes. MWGSP is by definition more versatile and is our main focus. 

Classical signal processing and GSP typically process one or two-dimensional signals~\cite{ortega2018graph,grassi2017time,romero2017kernel,Sandryhaila2014} and do not address datasets of higher dimensions. 
However, such datasets, given as multi-way tensors, are becoming increasingly common in many domains.
Mathematically, tensors generalize matrices to higher dimensions~\cite{KoldaBader2009}, and in this work the term tensors includes matrices (as they are 2-tensors). 
Examples of tensors includes video, hyperspectral imaging, MRI scans, multi-subject fMRI data, chemometrics, epigenetics, trial-based neural data, and higher-order \emph{sparse} tensor data such as databases of crime incident reports, taxi rides or ad click data~\cite{Mishne2016,Chi2018,sun2018convolutional,nie2017graph,zhang2018spatial, shahid2019tensor}.
While tensors are the primary structure for representing $D$-dimensional signals, research on tensors and signal processing on tensors has primarily focused on  factorization methods~\cite{KoldaBader2009,Cichocki2015}, devoting less attention to leveraging the  underlying geometry on the tensor modes. 
Recent MWGSP approaches incorporate graph smoothness in multiway tensor analysis, both for robust tensor factorization~\cite{nie2017graph,zhang2018spatial, shahid2019tensor} and direct data analysis of tensors~\cite{Mishne2016,Chi2018}. 




In this overview of {multi-way data} analysis, we present a broad viewpoint to simultaneously consider general graphs underlying all modes of a tensor. 
Thus, we interpret multi-way analyses in light of graph-based signal processing to consider tensors as \emph{multi-way graph signals} defined on multi-way graphs. GSP is a powerful framework in the multi-way setting, leading to intuitive and uniform interpretations of operations on irregular geometry. 
Thus, MWGSP is a non-trivial departure from classical signal processing, producing an opportunity to exploit joint structures and correlations across modes to 
more accurately model and process signals in real-world applications of current societal importance: climate, spread of epidemics and traffic, as well as complex systems in biology. 

Both the GSP and tensor analysis communities have been developing methods for multiway data analysis and have taken different but complementary strategies to solving common problems. 
We lay the mathematical and theoretical foundations drawing on work from both communities to develop a framework for higher-order signal processing of tensor data, and explore the challenges and algorithms that result when one imposes relational structure along all axes of data tensors.
At the heart of this framework is the graph Laplacian, which provides a basis for harmonic analysis of data in MWGSP and an important regularizer in modeling and recovery of multi-way graph signals. 
We illustrate the breadth of MWGSP by reinterpreting classic techniques, such as the 2-D discrete Fourier transform, as a special case of MWGSP and introduce a general Multi-Way Graph Fourier Transform (MWGFT). 
Further, we review novel multi-way regularizations that are not immediately obvious by viewing the data purely as a tensor.
Thus, we synthesize into a coherent family a spectrum of recent and novel MWGSP methods across varied applications in inpainting, denoising, data completion, factor analysis, dictionary learning, and graph learning~\cite{Chi2018, sun2018convolutional,rao2015collaborative, yankelevsky2016dual,li2015multi, ioannidis2018coupled,su2018graph,xie2018graph,rabin2019two}. 

The organization of this paper is as follows. Sec.~\ref{sec:gsp} reviews standard GSP, which we refer to as single-way GSP. 
Sec.~\ref{sec:mwgsp} introduces tensors and multilinear operators and constructs multi-way graphs, transforms, and filters. 
Sec.~\ref{sec:unique} briefly highlights two recent multiway frameworks: the time-vertex framework, a natural development of MWGSP that couples a known time axis to a graph topology, and the Generalized Graph Signal Processing framework which extends MWGSP by coupling non-discrete and arbitrary geometries into a single signal processing framework
Sec.~\ref{sec:mwsp} moves to multi-way signal modeling and recovery, where graph-based multi-way methods are used in a broad range of tasks. 
Sec.~\ref{sec:discussion} concludes with open questions for future work.

%% file: GSP.tex

GSP generalizes classical signal processing from \emph{regular} Euclidean geometries such as time and space, to irregular, and non-Euclidean geometries represented discretely by a graph. In this section, we review basic concepts.\footnote{A complete survey of graph signal processing is provided in \cite{ortega2018graph}.}

\paragraph{Graphs}This tutorial considers undirected, connected, and weighted graphs $\G = \{\Vc,\Ec,\M{W}\}$ consisting of a finite vertex set $\Vc$, an edge set $\Ec$, and a weighted adjacency matrix $\M{W}$. If two vertices $v_i, v_j$ are connected, then $(v_i,v_j)\in\Ec$, and $\M{W}_{i,j}= \M{W}_{j,i}>0$; otherwise  $\M{W}_{i,j}= \M{W}_{j,i}=0.$ We employ a superscript parenthetical index to reference graphs and their accompanying characteristics from a set of graphs $\G^{(i)}$, i.e., $G = \left\{\G^{(i)} = \left(\Vc^{(i)}, \Ec^{(i)},\M{W}^{(i)}\right)\right\}_{i=1}^D$. Contextually we will refer to the cardinality of the vertex set of a graph $\G^{(i)}$ as $\left|\Vc^{(i)}\right| = n_i$. When parenthetical indexing is not used, we refer to a general graph $\G$ on $n$ nodes. 
For details on how to construct a graph see the box ``Graph construction." 



\paragraph{Graph Signals} A signal $f: \Vc \rightarrow \Real ^n$ on the vertices of a graph on $n$ nodes may be represented as a vector $\mathbf{f} \in \Real^n$, where $\mathbf{f}_i={f}(i)$ is the signal value at vertex $v_i \in \Vc$.
The graph Fourier transform decomposes a graph signal in terms of the eigenvectors of a graph shift operator. Many choices have been proposed for graph shifts, including the adjacency matrix $\M{W}$ and various forms of the graph Laplacian $\M{\mathcal{L}}$, a second order difference operator over the edge set of the graph. 
In this paper we use the popular combinatorial graph Laplacian defined as $\M{\mathcal{L}}:=\mathbf{D}-\mathbf{W}$, where the degree matrix $\mathbf{D}$ is diagonal with elements $\mathbf{D}_{ii} = \sum_j \mathbf{W}_{ij}$. 
This matrix is real and symmetric. Its eigendecomposition is $\M{\mathcal{L}}=\M{\Psi}\mathbf{\Lambda}_\G\M{\Psi}^*$, where the columns of $\M{\Psi}$ are a complete set of orthonormal eigenvectors $\left\{\V{\psi}_{\ell}\right\}_{\ell=0}^{n-1}$, $\M{\Psi}^*$ is the conjugate transpose of $\M{\Psi}$, and the diagonal of $\mathbf{\Lambda}_\G$ are the real eigenvalues $\left\{\lambda_\ell \right\}_{\ell=0}^{n-1}$.

\paragraph{Graph Fourier Analysis}The Graph Fourier Transform (GFT) and its inverse are
\begin{equation}
\label{eq:gft}
\hat{f}(\lambda_{\ell}) := \langle {\mathbf{f},\V{\psi}_{\ell}}\rangle = \sum_{k=1}^N f(k) \psi_{\ell}^*(k) \quad\quad \text{and} \quad\quad f(k) = \sum_{\ell=0}^{N-1} \hat{f}(\lambda_{\ell}) \psi_{\ell}(k),
\end{equation}
or in matrix form $\text{GFT}\{\mathbf{f}\}=\M{\Psi}^*\mathbf{f}$.
The GFT generalizes the classical Fourier transform since the former is the spectral expansion of a vector in the discrete graph Laplacian eigensystem while the latter is the spectral expansion of a function in the eigensystem of the continuous Laplacian operator. Indeed, the GFT is synonymous with the discrete Fourier transform (DFT) when the graph Laplacian is built on a cyclic path or ring graph. 
It is typical to reinforce the classical Fourier analogy by referring to the eigenvectors of $\mathcal{L}$ as graph harmonics and the eigenvalues as graph frequencies and indexing the harmonics in ascending order of the eigenvalues such that the lowest indexed harmonics are the smoothest elements of the graph eigenbasis.



Despite these analogies, it is non-trivial to directly extend classical tools to signals on graphs. For example, there is no straightforward analogue of convolution in the time domain to convolution in the vertex domain. Instead, filtering signals in the GFT domain is defined analogously to filtering in the frequency domain, with a filtering function $\hat{h}(\cdot)$ applied to the eigenvalues $\lambda_\ell$, that take the place of the frequencies:
 \begin{equation}
\widetilde{f}(k)= \sum_{\ell=0}^{N-1} \hat{h}(\lambda_{\ell}) \hat{f}(\lambda_{\ell}) \psi_{\ell}(k),
\end{equation} where $\widetilde{f}$ is the result of filtering $f$ with the graph spectral filter
 $h(\mathcal{L})$. This spectral analogy is a common approach for generalizing classical notions that lack clear vertex interpretations.

%% file: gsp_tensors2.tex
Classical $D$-dimensional Fourier analysis provides a template for constructing unified geometries from various data sources. The $D$-dimensional Fourier transform applies a 1-dimensional Fourier transform to each axis of the data sequentially. For example, a 2D-DFT applied to an $n_1 \times n_2$ real image $\M{X}$ is 
\begin{equation}
\label{eq:dft2d}
    \text{2D-DFT}\{\M{X}\}=\mathrm{DFT}_c(\mathrm{DFT}_r(\mathbf{X})) = \mathrm{DFT}_r(\mathrm{DFT}_c(\mathbf{X}))=\M{U}_{n_1}\M{X}\M{U}_{n_2},
\end{equation} 

where $\mathrm{DFT}_{r}$ ($\mathrm{DFT}_{c}$) applies the DFT to the rows (columns) of $\mathbf{X}$ 
and $\M{U}_{n}$ denotes a normalized $n$-point DFT matrix: $\M{U}_n(t,k)=\frac{1}{\sqrt{n}} \exp\{-2\pi \mathbf{j}t(k-1)/n\} $ for $t,k = 1,\ldots, n$. 
This 2D transform decomposes the input into a set of plane waves.


The 2D graph Fourier transform (2D-GFT) is algebraically analogous to the 2D-DFT. For two graphs $\G^{(1)}$ and $\G^{(2)}$ on $n_1$ and $n_2$ vertices, the 2D-DFT  of $\M{X}\in \Real^{n_1\times n_2}$ is
\begin{equation}
\label{eq:2dgft}
\text{2D-GFT}(\M{X}) =\mathrm{GFT}_{n_1}\left(\mathrm{GFT}_{n_2}(\M{X}))\right)=\mathrm{GFT}_{n_2}\left(\mathrm{GFT}_{n_1}(\M{X})\right),
\end{equation}
and was presented in~\cite{Sandryhaila2014} as a method for efficiently processing big-data. 
Note that when $\G^{(1)}=\mathcal{P}^{n_1}$ and $\G^{(2)}= \mathcal{P}^{n_2}$, i.e., they are \emph{cyclic path graphs} on $n_1$ and $n_2$ vertices, this transform is equivalent to a 2D-DFT~\cite{Sandryhaila2014}. 

In this section, we present the MWGSP framework for general $D$-dimensional signal processing on coupled and irregular domains, which  enables holistic data analysis by considering relational structures on potentially {\em all} modes of a mutli-way signal. MWGSP encompasses standard GSP while extending fundamental GSP tools such as graph filters to $D$-dimensions. Furthermore, because graphs can be used to model discrete structures from classical signal processing, MWGSP forms an intuitive superset of discrete signal processing on domains such as images or video. 
\vspace{-1cm}
\subsection{Tensors}
\emph{Tensors} are both a data structure representing $D$-dimensional signals, 
as well as a mathematical tool for analyzing multilinear spaces. We use both perspectives to formulate MWGSP.
In this paper, we adopt the tensor terminology and notation used by \cite{KoldaBader2009}.
\subsubsection{Tensors as a D-dimensional array}
The number of ways or modes of a tensor is its \emph{order}. 
Vectors are tensors of order one and denoted by boldface lowercase letters, e.g.,\@ $\V{a}$. 
Matrices are tensors of order two and denoted by boldface capital letters, e.g.,\@ $\M{A}$. 
Tensors of higher-order, namely order three and greater, we denote by boldface Euler script letters, e.g.,\@ $\T{A}$. If $\T{A}$ is a $D$-way data array of size $n_1 \times \cdots \times n_D$, we say $\T{A}$ is a tensor of order $D$.

\begin{mdframed}[userdefinedwidth=\linewidth,align=center,usetwoside=false,rightmargin=-0.2in,
linecolor=blue,linewidth=1pt,frametitle = {Order-3 tensors}]
\small
For simplicity, we briefly review some tensor terminology for the $3$-way tensor $\T{X}\in \mathbb{C}^{n_1 \times n_2\times n_3}$. 
The size of each mode is denoted by $n_k$, with $n_1$ being the number of columns, $n_2$ the number of rows, and $n_3$ being the number of \emph{tubes}~\cite{KoldaBader2009}. Video and time-series recording of matrix valued signals are a common application for tensors of this form (\cref{fig:tensor}). In videos, the first and second modes of the tensor encode pixel values for each frame, while the third mode indexes the frames in time. 
We can \emph{slice} a video tensor to produce different views of the data as presented in Fig.~\ref{fig:tensor}. 

    \centering
    \includegraphics[width=0.95\linewidth]{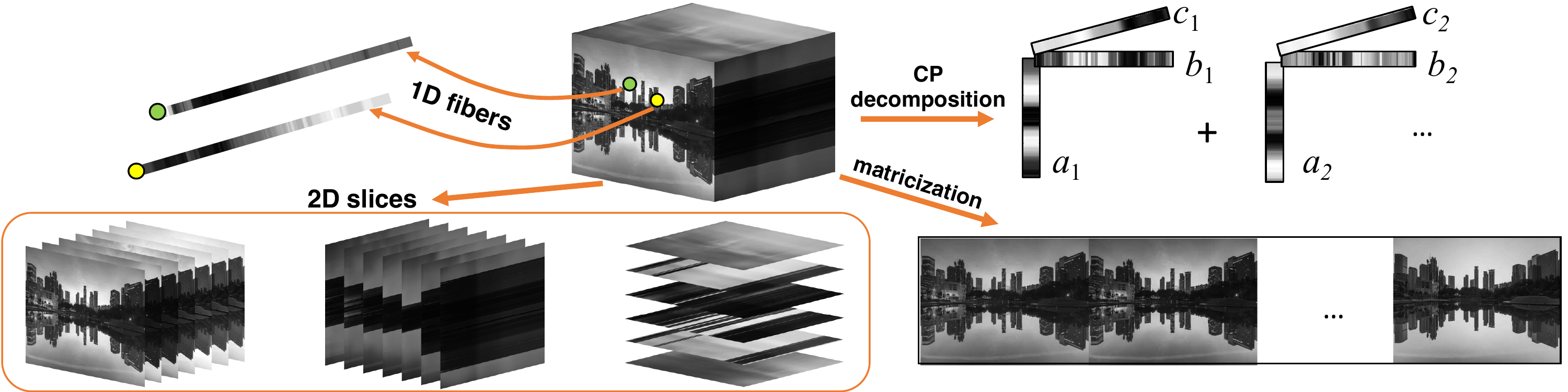}
    \captionof{figure}{\small Tensor terminology. A Time-lapse video is an order-3 tensor. Tensor slices (left to right): A \emph{frontal slice} is the matrix $\T{X}_{::k}$, formed by selecting the $k$-th frame of the video. The \emph{lateral slice}, $\T{X}_{:j:}$, is a matrix (viewable as an image) that shows the time evolution of the $j$-th column of pixels in the input. The \emph{horizontal slice} $\T{X}_{i::}$ similarly contains the time evolution of one row of pixels. 2D indexing of 3rd order tensors yields a \emph{1D fiber}. For example, the tubular fiber $\T{X}_{ij:}$ is an $n_3$ dimensional time-series of the $i,j$th pixel across all frames; the two tubular fibers correspond to the highlighted pixels in the tensor. Mode-1 \emph{matricization} concatenates all frontal slices side by side. \emph{CP decomposition} is a sum of rank-1 tensors.}
    \label{fig:tensor}
    \vspace{-0.2cm}
    
\end{mdframed}

There are multiple operations to reshape tensors, used for convenient calculations. \emph{Vectorization} maps the elements of a matrix into a vector in column-major order. That is, for $\M{X}\in \Real^{n_1\times n_2}$, $$\vec(\M{X}) = \left[ \M{X}_{1, 1}, \ldots,  \M{X}_{n_1, 1}, \M{X}_{1 2}, \ldots, \M{X}_{n_1 ,2}, \ldots,  \M{X}_{1 ,n_2}, \ldots,  \M{X}_{n_1, n_2}\right]^\mathrm{T}.$$ A tensor mode-$d$ vectorization operator, $\vec_d(\T{X})$ is similarly defined by stacking the elements of $\T{X}$ in mode-$d$ major order. 
Let $\mathrm{ten}\left(\V{x},\ell,\{n_1,\ldots,n_D\}\right) = \T{X}$ be the $\ell$-th \emph{tensorization} of $\V{x}$, which is the inverse of the $\ell$-major vectorization of $\T{X}$.
Denote by $n\setminus \ell = \prod_{i=1}^{\ell-1} n_i \prod_{j=\ell+1}^D n_j$ the product of all factor sizes except for the $\ell$-th factor. 
 Then, let $\mathrm{mat}\left(\T{X},\ell\right) = \M{X}^{(\ell)} \in \Real^{n_\ell \times n\setminus \ell}$ be the mode-$\ell$ matricization of $\T{X}$ formed by setting the $\ell$-th mode of $\T{X}$ to the rows of $\M{X}^{(\ell)}$, vectorizing the remaining modes to form the columns of $\M{X}^{(\ell)}$ as in Fig.~\ref{fig:tensor}. 



\subsubsection{Tensor products}
Up to this point we have avoided explicitly constructing $D$-dimensional transforms.
In the 2D case, applying a two-dimensional transform is calculated via linear operators as in~(\ref{eq:dft2d});  generalizing to higher-order tensors requires multilinear operators.   
Therefore, we introduce the \emph{tensor product} and its discrete form, the \emph{Kronecker product}. These products are powerful tools for succinctly describing $D$-dimensional transforms.

The great utility of the tensor product is that it simultaneously transforms spaces alongside their linear operators. This is the so-called \emph{universal property} of the tensor product.  In brief, it states that the tensor product, denoted by $\otimes$, of two vector spaces $V$ and $W$ is the unique result of a bilinear map $\varphi : V \times W \rightarrow V \otimes W$.  The power in $\varphi$ is that it uniquely factors any bilinear map on $V\times W$ into a linear map on $V \otimes W$. The universal property implies that the tensor product is symmetric: $V \otimes W$ is a canonical isomorphism of $W \otimes V$.  Though the tensor product is defined in terms of two vector spaces, it can be applied repeatedly to combine many domains, so we generically refer to it as a product of many spaces.

In this paper, we are concerned with the tensor product on Hilbert spaces $\mathcal{H}^{(k)},~ k=1,\ldots, D$. These metric spaces include both continuous and discrete Euclidean domains from classical signal processing, as well as the non-Euclidean vertex domain. Since tensor products on Hilbert spaces produce Hilbert spaces, we can combine time, space, vertex, or other signal processing domains via the tensor product and remain in a Hilbert space. Under some constraints, an orthonormal basis for the product of $D$ Hilbert spaces is admitted directly by the tensor product of the factor spaces. These properties of the tensor product are the mathematical foundations for the remainder of this tutorial, in which we construct a multi-way signal processing framework based on unifying multiple input spaces and their Fourier operators into a single linear representation.

\paragraph*{Kronecker products}
The Kronecker product produces the matrix of a tensor product with respect to a standard basis and generalizes the outer product of vectors $\V{x}\V{y}^*$ for $\V{x}\in\Complex^{m}$ and $\V{y} \in \Complex^{n}$. For analogy, it is common to use the same notation to denote the Kronecker and tensor product.
The Kronecker product is associative. Consequently the matrix $\M{M}$ that is the Kronecker product  of a sequence of $D$ matrices $\M{M}^{(k)} \in \Complex^{n_k \times n_k}$ for $k=1,\ldots,D$ is 
\begin{equation}
\begin{split}
\M{M} = \bigotimes_{k=1}^D \M{M}^{(k)} =  \M{M}^{(1)} \Kron \left(\bigotimes_{k=2}^D \M{M}^{(k)}\right) = \left(\bigotimes_{k=1}^{D-1} \M{M}^{(k)}\right)\Kron \M{M}^{(D)} = \M{M}^{(1)}\Kron \cdots \Kron \M{M}^{(D)}.
\label{eq:general_kronecker_product}
\end{split}
\end{equation}
It is important to note that the Kronecker product is in general non-commutative. For brevity, we will apply a decremental Kronecker product using the notation 

$$ \downarrow\bigotimes_{k=1}^D \M{M}^{(k)} = \bigotimes_{k=0}^{D-1} \M{M}^{(D-k)} =
\M{M}^{(D)}\Kron \cdots \Kron \M{M}^{(1)}.$$

The Kronecker product has many convenient algebraic properties for computing multidimensional transforms.
Vectorization enables one to express bilinear matrix multiplication as a linear transformation
\begin{equation}
    \vec\left(\M{C}\M{X}\M{B}\right)=\left(\M{B}\Tra\Kron\M{C}\right)\vec{\left(\M{X}\right)},
    \label{eq:vectrick} \vspace{-0.2cm}
\end{equation}
assuming that the dimensions of $\M{C},\M{X},\M{B}$ are compatible such that $\M{C}\M{X}\M{B}$ is a valid operation.
This identity is a discrete realization of the universal property of tensors, and shows that the Kronecker product corresponds to a bilinear operator. We will use this identity to \begin{enumerate*} \item construct multi-dimensional discrete Fourier bases, and \item  decompose multi-way algorithms for computational efficiency. \end{enumerate*} 
\subsection{Multi-way transforms and filters}
We now apply~(\ref{eq:vectrick}) to explicitly construct a 2D-GFT.
 If $\M{\M{\Psi}}^{(1)}$ and $\M{\M{\Psi}}^{(2)}$ are Fourier bases for graph signals on any two graphs $\G_1$ and $\G_2$, a 2D-GFT basis is $\M{\M{\Psi}}^{(2)} \Kron \M{\M{\Psi}}^{(1)}$. This is a single orthonormal basis of dimension $\left|\Vc^{(1)}\right|\left|\Vc^{(2)}\right|\times \left|\Vc^{(1)}\right|\left|\Vc^{(2)}\right|$, which can be used to describe a 2D graph signal $\M{X}\in\Real^{n_1\times n_2}$ in the geometry of a single multi-way graph by the GFT $$\hat{\V{x}}=\left(\M{\M{\Psi}}^{(2)}\Kron\M{\M{\Psi}}^{(1)}\right)^* \vec\left(\M{X}\right).  \vspace{-0.2cm}$$ 
 
 Unlike the DFT, where it is clear that increasing dimension yields grids, cubes, and hypercubes, interpreting the geometry of $\M{\M{\Psi}}^{(2)} \Kron \M{\M{\Psi}}^{(1)}$ is less intuitive. For this, we must turn to a \emph{graph product}.


\paragraph*{Product graphs} MWGSP relies on a graph underlying each mode of the given tensor data. The question is: What joint geometry arises from these graphs, and what multilinear operators exist on this joint graph structure? Our approach is to construct a multiway graph $\G= \{\Vc,\Ec,\M{W}\}$ over the entirety of a data $\T{X}$ as the \emph{product graph} of a set of \emph{factor} graphs $G = \left\{\G^{(1)},\ldots, \G^{(D)}\right\}$. 

For example, if $\T{X}\in\Real^{n_1\times n_2\times n_3}$ contains the results of a $n_3$ sample longitudinal survey of $n_2$ genes on a cohort of $n_1$ patients, then the intramodal relationships of $\T{X}$ are modeled by separate graphs $\G^{(1)}$, in which each patient is a vertex, $\G^{(2)}$, in which each gene is a vertex, and $\G^{(3)}$, which represents time as a path graph on $n_3$ vertices. We will use this example throughout this section, though our derivation generalizes to tensors of arbitrary order.

While one could treat matrix-valued slices of $\T{X}$ as signals on each individual graph, we use the graph product to model $\T{X}$ as a single graph signal on $\G$. We begin by constructing $\Vc$, the vertices of $\G$, which for all graph products is performed by assigning a single vertex to every element in the Cartesian product of the factor vertex sets  i.e., $\Vc = \Vc^{(1)} \Cart \ldots \Cart \Vc^{(D)}$. Thus, the cardinality of the vertex set of $\G$ is $n=\prod_{k=1}^D n_k$. For example, our longitudinal survey will be modeled by the product graph $\G$ on $n = n_1n_2n_3$ vertices. As a Cartesian product, the elements $v\in\Vc$ can be expressed as the tuple $v=$(\emph{patient}, \emph{gene}, \emph{time}). The experimental observation tensor can be modeled as a graph signal\footnote{We can do this because the vectorization $\vec({\T{X}})$ is isomorphic to $\T{X}$, which can be shown using~(\ref{eq:vectrick}).} $\V{x}=\vec(\T{X})$ in $\Real^{n}$.

 Our next step is to learn the topology of $\G$ by mapping the edge sets (weights) of the factor graphs into a single set of product edges (weights) $\Ec$. There are a variety of graph products, each of which differs from each other only in the construction of this map. We focus on the \emph{Cartesian graph product} as it is the most widely employed in multi-way algorithms. However, other products such as the tensor and strong graph products each induce novel edge topologies that warrant further exploration for MWGSP \citep{Sandryhaila2014}.

\begin{figure*}
     \centering
    \includegraphics[width=0.9\linewidth]{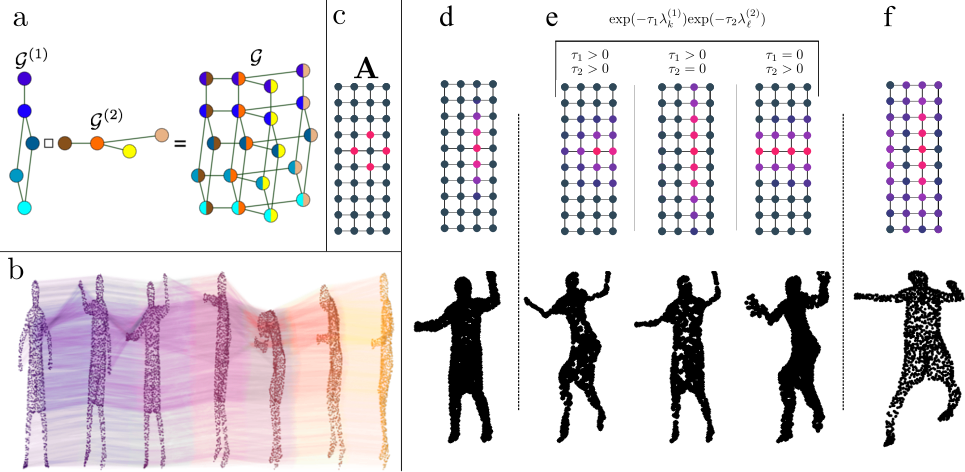}
    \captionof{figure}{\small Multi-way graphs, signals and spectral filters. (a) The Cartesian graph product generates a copy of $\G^{(1)}$ at each vertex of $\G^{(2)}$. (b) A multiway graph formed from a dynamic mesh~\cite{grassi2017time}. This time-vertex graph (purple = $t_0$, yellow = $t_7$) connects each point in the mesh to its counterpart in adjacent frames. The temporal evolution of the 3D coordinates is a graph signal on this graph. (c) A single column of the joint adjacency matrix of a 2D grid shifts signals to their neighbors. (d-f) Multiway filtering. (Top) A multiway signal on a 2D grid. This signal can be decomposed into an impulse and a smooth signal; thus it is bandlimited along one way of the grid.  (Bottom) A frame of the Dancer mesh. (e) A separable diffusion filter is the product of domain-specific heat kernels. Separable filters can filter along both axes in unison~(left), or each axis independently (middle / right). (f) Non-separable filter. For the dynamic mesh, filtering along only one mode reveals either skeleton structure ($\tau_2=0$) or averaged (blurred) dynamics of the figure ($\tau_1=0$), but joint separable or non-separable filtering reveals joint dependencies. }
    \label{fig:product_graphs}
        \vspace{-0.3cm}
 \end{figure*}

\paragraph*{Cartesian graph products}
We denote the Cartesian product of $D$ graphs as
\begin{align}
\G = \GCart\limits_{\ell=1}^D \G^{(\ell)} = \G^{(1)}\Gcart\cdots\Gcart\G^{(D)}.
\end{align}
The Cartesian graph product is intuitively an XOR product since for any two vertices 
\begin{align}\left\{v=\left(v^{(1)},\ldots,v^{(D)}\right),u=\left(u^{(1)},\ldots,u^{(D)}\right)\right\}\subset \Vc,
\end{align}
the edge $(v,u)$ 
exists if and only if there exists a single $i$ such that $(v^{(i)},u^{(i)}) \in \Ec^{(i)}$ and $v^{(\ell)} = u^{(\ell)}$ for all $\ell\neq i$. In other words, the vertices of $\G$ are connected if and only if exclusively one pair of factor vertices are adjacent and the remaining factor vertices are the same.~\Cref{fig:product_graphs}a illustrates the generation of an $n_1\times n_2$ 2D grid graph via the product of two path graphs on $n_1$ and $n_2$ vertices. 

The Cartesian graph product can induce topological properties such as regularity onto a graph. Since the path graph basis is well-characterized as a discrete Fourier basis, it is a convenient tool for including Euclidean domains in multi-way analysis.  For example, we can model time series and longitudinal graph signals as a single vector using a path graph product. In the case of our gene expression data $\T{X}$, the product of the gene and patient mode graphs with a path on $n_3$ vertices, i.e., $\G^{(1)}\Gcart\G^{(2)}\Gcart \mathcal{P}^{n_3}$, models the data by treating the temporal mode as a sequence. One can intuit this operation as copying $\G^{(1)}\Gcart\G^{(2)}$ $n_3$ times and connecting edges between each copy.

\paragraph*{Product graph matrices}
The Kronecker product links graph shift operators on Cartesian product graphs to the corresponding operators on the factors. The \emph{Kronecker sum} of $D$ matrices $\M{A}^{(k)} \in \Complex^{n_k\times n_k}$ for $k=1,\ldots,D$ is 

 \begin{equation}
 \M{A}= \bigoplus_{k=1}^D \M{A}^{(k)} = \sum_{k=1}^{D} \M{I}_{n_{>k}}\Kron{}\M{A}^{(k)}\Kron{}\M{I}_{n_{<k}},\label{eq:general_kronecker_sum} \text{where} ~ n_{>k} = \prod_{\ell=k+1}^D n_\ell, ~ \text{and}~ n_{<k} = \prod_{\ell=1}^{k-1} n_\ell\nonumber.
 \end{equation}
 The joint adjacency matrix $\M{A}$ and graph Laplacian $\M{\mathcal{L}}$ are constructed by the Kronecker sum of their corresponding factor graph matrices.
~The eigensystem of a Kronecker sum is generated by the pairwise sum of the eigenvalues of its factors and the tensor product of the factor eigenbases  \citep[][Thm. 4.4.5]{horn1994topics}. Thus, the Fourier basis $\Psi$ for the product graph $\G$ is immediate from the factors.
For $k=1,\ldots,D$~let $\left(\lambda_{\ell_k},\V{\psi}_{\ell_k}\right)$ be the $\ell_k$th eigenpair of $\M{\mathcal{L}}^{(k)}$ for $0\leq\ell_k\leq n_k-1$
Then let $I_{\ell} = (\ell_1,\ldots,\ell_D) \in [n_1]\times\ldots \times [n_D]$
be a multi-index to the $\ell$th eigenpair of $\M{\mathcal{L}}$.  The product graph Fourier basis is then
\begin{align}
\label{eq:general_kron_sum_eigensystem}
\left(\lambda_{I_{\ell}},\V{\psi}^{}_{I_{\ell}}\right) = \left(\sum_{k=1}^D \lambda^{(k)}_{\ell_k}, \downarrow\bigotimes_{k=1}^D \V{\psi}^{(k)}_{\ell_k}\right).
\end{align} 

Thus, the MWGFT of a multiway graph signal $\T{X}$ is  
\begin{align}
\V{\hat{x}}= \M{\Psi}^*\vec(\T{X})=\left(\downarrow\bigotimes_{k=1}^{D} \M{\Psi}^{(k)} \right)^\ast\vec(\T{X})
.
\label{eq:directapplication}
\end{align} 
This formulation includes applying a single-way transform along one mode of the tensor, for example, DFT$_{n_1}\{\T{X}\}=\left(\downarrow \bigotimes_{k=2}^{D} \M{I}_{n_k} \Kron \M{U}_{n_1} \right)^T\vec(\T{X})$ applies the DFT along the first mode of the tensor.

\paragraph*{Efficient MWGSP by graph factorization}
On the surface, the computational cost of a MWGFT (and MWGSP in general) seems high as 
multi-way product graphs are often much larger than their individual factors; the cardinality of the product vertex set is the product of the number of vertices in each factor.  However, the product graph structure actually yields efficient algorithms. 
With small adjustments to fundamental operations like matrix multiplication, in the best case one can effectively reduce the computational burden of an order $D$ tensor with $n = \prod_{\ell=1}^Dn_\ell$ total elements to a sequence of problems on $n^{(1/D)}$ elements. The computational strategy is to apply~\Cref{eq:vectrick} and its order-$D$ generalization to avoid storing and computing large product graph operators. 

We introduce the order-$D$ form of~\Cref{eq:vectrick} via an algorithm. Given a sequence of operators $\M{M}^{(\ell)}\in \Real^{n_\ell},~\ell=1,\ldots, D$, an efficient algorithm for computing 
$
\V{y}= \bigotimes_{\ell=1}^D \M{M}^{(\ell)}\vec(\T{X})
$
proceeds by applying each $\M{M}^{(\ell)}$ to the corresponding mode-wise matricization of $\T{X}$. \Cref{alg:mode-D} presents pseudocode for computing this product.
\begin{algorithm}
\caption{$D$-tensor multilinear transformations}
\begin{algorithmic}[1]
\State Initialize $\T{Y} =\T{X}$
\For{$\ell=1,\ldots,D$}
\State Matricize: $\M{Y}^{(\ell)} = \mathrm{mat}\left(\T{Y},\ell\right)$
\State Factor update: $\M{Y}^{(\ell)} = \M{M}^{(\ell)\Tra} \M{Y}^{(\ell)}$
\State  Reform tensor: $\T{Y} = \mathrm{ten}\left(\M{Y}^{(\ell)},\ell,\{n_1,\ldots,n_D\}\right)$
\EndFor
\State Vectorization: $\V{y} = \mathrm{vec}\left(\T{Y}\right).$
\end{algorithmic}
\label{alg:mode-D}
\end{algorithm}

As a sequential product of an $n_\ell\times n_\ell$ matrix with an $n_\ell \times n \setminus \ell$ matrix, this method can dramatically improve the cost of algorithms that depends on matrix multiplication. Further, the number of operations only depends on computations over smaller factor matrices, enabling one to perform computations on the product graph without computing and storing expensive operators. 

For example, consider the computational cost of applying an MWGFT for a product graph $\G$ on $n=\prod_{\ell=1}^D n_\ell$ nodes. In the worst case, \Cref{alg:mode-D} is as fast as directly computing~(\ref{eq:directapplication}). However, in the best-case scenario  $n_\ell=\sqrt[D]{n}$ for all $\ell=1,\ldots,D$, and computing $D$ graph Fourier bases of size $\sqrt[D]{n} \times \sqrt[D]{n} $ requires $\mathcal{O}\left(n^{3/D}\right)$ operations. To compute a MWGFT using the factor bases, we use $\Psi^{(\ell)}$ as the sequence of operators in Alg.~\ref{alg:mode-D}, which costs  $\mathcal{O}\left(D{n}^{1/D+1}\right)$ operations. This improves upon the standard GFT, which costs $\mathcal{O}\left(n^3\right)$ operations to obtain an eigenbasis and $\mathcal{O}\left(n^2\right)$ operations to apply. For example, when $D=3$ and $n_1=n_2=n_3=\sqrt[3]{n},$ we obtain an asymptotically linear factorization of a graph Fourier basis for $\G$, and the corresponding MWGFT can be applied in $\mathcal{O}\left(3{n}^{1/3+1}\right)$ operations.

\begin{mdframed}[userdefinedwidth=\linewidth,align=center,usetwoside=false,rightmargin=-0.2in,
linecolor=blue,linewidth=1pt,frametitle = {Multi-way signal compression}]
\small
A key motivation for MWGSP is the capability of encoding multi-way data in a compact way. Transforms with good energy compactness summarize the data well and can be used to construct efficient regularizers for regression problems. Figure~\ref{fig:compression} demonstrates energy compression in four datasets. The Dancer mesh~\cite{grassi2017time} shown in Fig.~\ref{fig:product_graphs}b couples $n_1 = 1502$ points to temporal evolution across $n_2 = 570$ timesteps. The Molene weather dataset~\cite{loukas2019stationary} ($n_1=32$ weather stations measuring temperatures over $n_2=24$ hours across $n_3=30$ days) couples a spatially determined graph to two temporal scales (hours, days). The time lapse video~\cite{grassi2017time} couples a 2D spatial grid ($492 \times 853$ pixels) to a temporal axis ($602$ timesteps), while the hyperspectral dataset~\cite{aviris_data} couples a 2D spatial grid ($145\times145$ pixels) to 200 spectral bandwidths (treated as a graph). All graphs were constructed using $k$-nearest neighbors with weighted edges set using a Gaussian kernel on the matricized modes of the tensor.
To measure energy compactness, we compute the relevant among the GFT, DFT (temporal axis), JFT, 2D-DFT (spatial grid), 3D-DFT (spatail grid+temporal axis) and MWGFT (all tensor modes) transforms for each dataset. We replace the spectrum coefficients with magnitudes smaller than the p-th percentile with zeros and perform the corresponding inverse transform on the resulting coefficients. 
The normalized compression error is computed from the signal reconstructed after thresholding the values of the transforms below the $p$-th percentile, denoted $\T{X}_p$, and given by $\Vert \textrm{vec}(\T{X}_p-\T{X}) \Vert_2 /\Vert \textrm{vec}(\T{X}) \Vert_2$. MWGFT achieves the best compactness in all datasets, providing insight that there are advantages to treating classical domains (time and space) as lying on graphs themselves.
\vspace{0.2cm}

     \centering
    \includegraphics[width=0.99\linewidth]{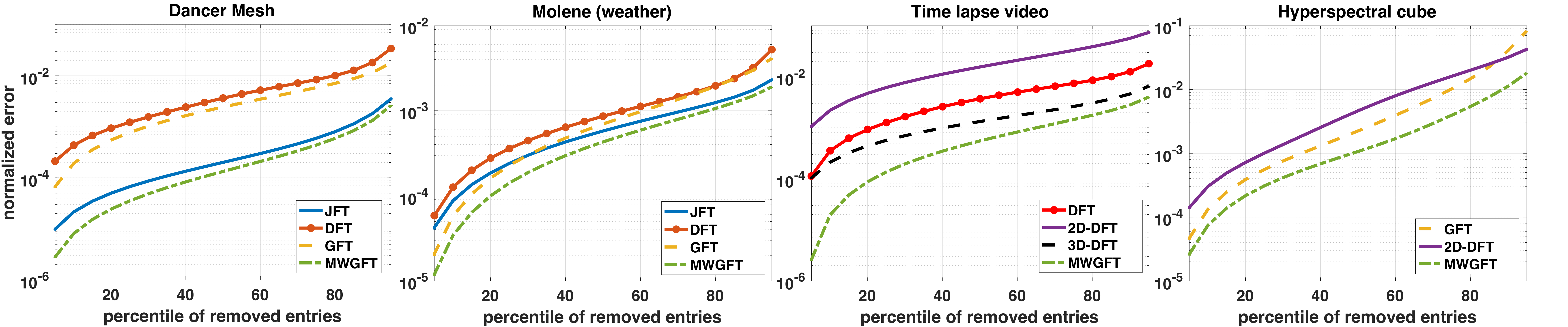}
    \captionof{figure}{\small Compactness of single-way and multi-way transforms for different datasets: dancer mesh (Fig.~\ref{fig:product_graphs}), Molene weather, time-lapse video (Fig.~\ref{fig:tensor}) and AVIRIS Indiana Pines hyperspectral image. }
    \label{fig:compression}
    \vspace{-0.2cm}

\end{mdframed}

\paragraph*{Edge density}
The graph edge density impacts the scalability of signal processing algorithms for multi-way data. Matrix equations can be efficiently solved by iteratively computing sparse matrix-vector products. The computational complexity of such algorithms, which include fundamental techniques such as Krylov subspace methods and polynomial approximation, typically depend linearly on the number of edges in the graphs, e.g.,\@ \cite{grassi2017time,Shahid2016, Chi2017a}. This dependency suggests using the sparsest possible graph that still captures the main similarity structure along each mode. Indeed, a common strategy is to construct sparse Laplacian matrices \cite{Shahid2016} or edge-incidence matrices \cite{Chi2017a}, using $k$-nearest-neighbor graphs which produce edge sets whose cardinality is linear in the number of nodes.
Yet, given sparse factors graph, there is no guarantee that the product will be sparse. Thus, major efficiency gains for multi-way algorithms can be made by replacing iterative matrix-vector multiplications (both sparse and dense) with a sequence of factor graph sparse matrix-vector multiplications using~\Cref{alg:mode-D}.

Three immediate applications for such a factorization are multi-way filter approximations~\citep[see, e.g.][]{grassi2017time}, compressive spectral clustering~\cite{tremblay2016compressive}, and fast graph Fourier transforms~\cite{frerix2019approximating}. We detail the former, while briefly describing future directions for the latter. For filtering, one could spectrally define and exactly compute a multi-way product graph filter (see box on Multi-way Filters) using the MWGSP techniques described in the previous section.  Yet, Chebyshev approximations~\cite{grassi2017time} are an efficient, robust, and accurate technique for approximate filtering. These approaches approximate spectrally defined filters by applying a recurrently defined weighted matrix-vector multiplication. Efficient multi-way Chebyshev approximation leverages the Kronecker sum definition for product graph Laplacians $\mathbf{\mathcal{L}}$. That is, by noting that $\mathbf{\mathcal{L}}\V{x} = \sum_{k=1}^D \left(\M{I}_{n_{>k}}\Kron{}\M{\mathcal{L}}^{(k)}\Kron{}\M{I}_{n_{<k}}\right)\V{x}$ is equivalent to computing $$\left(\M{I}_{n_{>1}}\Kron{}\M{\mathcal{L}}^{(1)}\right)\V{x} +\left( \M{I}_{n_{>2}}\Kron{}\M{\mathcal{L}}^{(2)}\Kron{}\M{I}_{n_{1}}\right)\V{x}+\ldots+\left(\M{\mathcal{L}}^{(D)}\M{I}_{n_{<D}}\right)\V{x},$$ it is clear that Chebyshev approximations of functions on $\mathbf{\mathcal{L}}$ (such as spectral graph wavelets) can be written as a sum of sparse matrix vector multiplications; the total operations are now dominated by the densest factor graph.

The efficiency of this approach cannot be understated, as it facilitates many algorithms, including the compressive spectral algorithm~\cite{tremblay2016compressive}. Indeed, it is increasingly common to estimate geometric and spectral qualities of the graph Laplacian by applying ideal filter approximations for eigencounting and coherence estimation.  Finally, factor graph sparsity and~\Cref{alg:mode-D} could be combined with recently proposed approaches for approximate orthogonal decompositions~\cite{frerix2019approximating} to construct a \emph{fast product graph Fourier transform}. This algorithm would admit striking similarities to the classical fast Fourier transform.

\begin{mdframed}[userdefinedwidth=\linewidth,align=center,usetwoside=false,rightmargin=-0.2in,
linecolor=blue,linewidth=1pt,frametitle = {Multi-way Filters}]
\label{sec:multi-way_filters}
\small{ 
It is natural to define spectral filters for multi-way graph signals on the product $\G$ as a function over the product graph eigenvalues $h: \Lambda \mapsto \Real{}$ as if they are traditional spectrally defined GSP filters.  Since these functions operate on the product eigenvalues, they directly consider the edge topology induced by a particular choice of product. 
Yet, it is feasible to develop filters for multi-way graph signals on $\G$ that are defined by multivariate functions $h : \Lambda^{(1)}\Cart\cdots\Cart\Lambda^D \mapsto \Real{}$. These multivariate filters are split into two classes: separable and nonseparable.

Separable filters have multivariate response functions that can be written as the product of separate univariate functions. In the $D=2$ case, a separable filter for the product graph could be written as $\left(\M{H}^{(2)}\Kron{}\M{H}^{(1)}\right)\V{x}$ in which $\M{H}^{(1)} = \M{\M{\Psi}}^{(1)} h^{(1)}(\M{\Lambda}^{(1)}) \M{\M{\Psi}}^{(1)^*}$ and $\M{H}^{(2)} = \M{\M{\Psi}}^{(2)} h^{(2)}(\M{\Lambda}^{(2)}) \M{\Psi}^{(2)^*}.$ Since this Kronecker product is permutation equivalent, we can treat its operation as an order-independent unimodal filtering of $\V{x}$~(\ref{eq:vectrick}). If $\M{H}^{(1)}$ and $\M{H}^{(2)}$ are both filters defined in a Laplacian eigenbasis of their respective factor graph, then the tensor product $\left(\M{H}^{(2)}\Kron{}\M{H}^{(1)}\right)$ is also diagonalized by the product eigenbasis. Thus, this filter is merely a reweighting of the product graph eigenbasis. 
 In~\Cref{fig:product_graphs}e, we demonstrate the application of a product of mode-wise heat filters to a graph signal on a grid~(\cref{fig:product_graphs}d top) and to a time-vertex signal which is a dynamic mesh~(\cref{fig:product_graphs}d bottom). While there is a choice of $\tau_1$ and $\tau_2$ such that certain regions of this filter can be computed from a heat kernel on the Cartesian product graph spectrum, such an approach abandons the flexibility of bilinear filtering. 
 By separability, each mode can be analyzed independently of the other by setting the appropriate $\tau_k$ to 0. This enables analyzing a joint signal along each mode independently, for example by filtering out high frequency structure along one domain while preserving the frequency content of the other mode.
A $D$-way separable filter applied to $\V{x} = \vec\left(\T{X}\right)$ is given by 
 \vspace{-0.2cm}
 \begin{align*}
 \V{\widetilde{x}}=\M{\Psi}h\left(\Bboxtimes_{k=1}^D\Lambda^{(k)}\right)\M{\Psi}^* \V{x},
 \end{align*}
  where 
  $h\left(\Bboxtimes_{k=1}^D\Lambda^{(k)}\right)$ is a diagonal matrix whose elements are given by $\prod_{k=1}^D h^{(k)}\left(\lambda_{\ell_k}\right)$, i.e., the product of separate spectral functions $h^{(k)}$ for each factor graph $\G^{(k)}$.

 Nonseparable filters cannot be designed from separate univariate filters on each mode. This class of filters encompasses a broad class of functions that include many filters defined in terms of the product graph eigenvalues, as well as multivariate functions (\cref{fig:product_graphs}f). Indeed,~\cite{grassi2017time} find that one cannot in general describe PDEs that describe diffusion, wave, or disease propagation with separable filters, as the relation between frequencies is not independent. 
 }
 \end{mdframed}


%% file: extensions.tex
\section{MWGSP frameworks}
\label{sec:unique}
Here we highlight two recent multi-way frameworks: time-vertex framework~\cite{grassi2017time} and Generalized GSP~\cite{ji2019hilbert}.

\vspace{-0.35cm}
\subsection{Time-vertex framework} The \emph{joint time-vertex} (T-V) framework~\cite{grassi2017time, romero2017kernel,loukas2019stationary} arose to address the limitations of GSP in analyzing dynamic data on graphs. 
This required generalizing harmonic analysis to a coupled time-graph setting by connecting a regular axis (time) to an arbitrary graph. The central application of these techniques are to analyze graph signals that are time-varying, for example, a time-series that reside on a sensor graph. Each time point of this series is itself a graph signal, while each vertex on the graph maps to a time-series of $T$ samples.  
This enables learning covariate structures from T-V signals, which are bivariate functions on the vertex and time domain. Such sequences of graph signals are commonly collected longitudinally through sensor networks, video, health data, and social networks.

The Joint time-vertex Fourier Transform (JFT)~\cite{grassi2017time} for a T-V signal $\M{X} \in \mathbb{R}^{|V| \times T}$  is defined as
$$\mathrm{JFT}\{\M{X}\}= \M{\Psi}^* \M{X}\M{U}_T \quad \text{or} \quad \mathrm{JFT}\{\vec(\M{X})\}= (\M{\bar{U}}_T\Kron \M{\Psi})^* \vec(\M{X}),$$
such that the multi-way Fourier transform of a T-V signal is a tensor product of the DFT basis with a GFT  basis (see Fig.~\ref{fig:compression}). Consequently, the JFT admits a fast transform in which one first performs an FFT along the time mode of the data before taking the GFT of the result, thus requiring only one Laplacian diagonalizaiton.

Including the DFT basis in this framework immediately admits novel joint time-vertex structures that are based on classical tools, such as variational norms that combine classical variation with graph variation~\cite{grassi2017time} introduce. For efficient filter analysis, they also propose an FFT and Chebyshev based algorithm for computing fast T-V filters, which applies to both separable and non-separable filters; see an example of T-V filtering in Fig.~\ref{fig:product_graphs}. 
Finally, overcomplete dictionary representations are constructed as a tensor-like composition of graph spectral dictionaries with classical short-time Fourier transform (STFT) and wavelet frames. These joint dictionaries can be constructed to form frames, enabling the analysis and manipulation of data in terms of time-frequency-vertex-frequency localized atoms. 
T-V spectral filtering was also introduced in~\cite{romero2017kernel}, as well as a T-V Kalman filter, with both batch and online function estimators.  
Further works have integrated ideas from classical signal processing such as stationarity to graph and T-V signals~\cite{girault2015stationary,marques2017stationary,loukas2019stationary}.
Thus, recent developments in the T-V framework can serve as a road-map for the future development of general MWGSP methods.
\vspace{-0.5cm}
\subsection{Generalized Graph Signal Processing}
Another recent development is that of the Generalized Graph Signal Processing~\cite{ji2019hilbert} framework which extends the notions of MWGSP to arbitrary, non-graphical geometries. Generalized GSP facilitates multivariate signal processing of interesting signals in which at least one domain lacks a discrete geometry. This framework recognizes that the key intuition of graph signal processing is the utility of irregular, non-Euclidean geometries for analyzing signals. However, where GSP techniques axiomatize a finite relational structure encoded by a graph shift operator, Generalized GSP extends classical Fourier analogies to arbitrary Hilbert spaces (i.e., complete inner product spaces) $\mathcal{H} \in H$ equipped with a compact, self-adjoint operator $A$. This broad class of geometries contains GSP, as the standard space of square summable graph signals, i.e., $L^2(V) = \{ f : V\mapsto  \Complex ~\text{,}~ \|{f}\|_2<\infty \}$ is itself a Hilbert space.

The geometries and corresponding signals that can be induced by Generalized GSP offer an intriguing juxtaposition of continuous and discrete topologies. As an example, consider the tensor product of a graph $\G$ with the space of square integrable functions on an interval, e.g.\@ $\G\Kron L^2([-1,1])$. Graph signals in this space map each vertex to a $L^2$ function. Conversely, $L^2$ functions can be mapped to specific vertices. To generate a Fourier basis for the product space, one simply takes the tensor product of the factor space eigenbases. This is a promising future direction for MWGSP, as it implies that one can, for instance, combine graph Fourier bases with generalized Fourier bases for innovative signal representations.

\cite{sun2018convolutional} proposed an early example of Generalized GSP, though under a different name. This work modeled videos and collections of related matrices as matrix-valued graph signals using \emph{matrix convolutional networks}. The authors aimed to solve the challenging missing data problem of node undersampling: some matrix slices from the networks are completely unobserved. When matrices have a low-rank graph Fourier transform, the network's graph structure enables recovery of missing slices. In light of the development of Generalized GSP, it is clear that~\cite{sun2018convolutional} proposed an algorithm for denoising of multi-way signals on $\G \Kron \Real^{n_1\times n_2}$.

%% file: graph_regularize.tex
In the previous section, we focused on signal processing through the lens of harmonic analysis, using the graph Laplacian to analyze data in the spectral domain.
In this section, we focus on signal modeling and recovery in the multi-way setting through the lens of optimization, where the graph Laplacian serves the role of imposing signal smoothness. 
Including graph structures along the modes of multi-way matrices and higher-order tensors has led to more robust and efficient approaches for denoising, matrix completion and inpainting, collaborative filtering, recommendation systems, biclustering, factorization, and dictionary learning~\cite{rao2015collaborative,yankelevsky2016dual,ioannidis2018coupled,sun2018convolutional,Chi2018,rabin2019two}. 
We begin with dual-graph modeling in the matrix setting and then extend to the higher-order tensor setting.
In the tensor setting we review both using multi-way graph regularization in tensor factorization methods and in complementary fashion using tensor factorization in signal modeling and recovering to make graph regularization computationally tractable. 
\vspace{-1cm}
\subsection{Signal processing on dual graphs}
The quadratic form of the graph Laplacian of a graph $\mathcal{G}$ quantifies the \emph{smoothness} of a signal $\V{f}$ with respect to the graph, where the smoother a signal is the smaller the value: 
\begin{equation}
\label{eq:quad}
    \mathbf{f}\Tra\M{\M{\mathcal{L}}}\mathbf{f} = \sum_{(i,j) \in\Ec}\M{W}_{i,j}\left(\mathbf{f}_i-\mathbf{f}_j\right)^2.
\end{equation}
Consequently, the typical model in th multi-way signal recovery setting is to add dual row-column graph regularizers of the form $\gamma_r\textrm{Tr}\left(\M{X}\Tra\M{\M{\mathcal{L}}}_r\M{X}\right)+\gamma_c \textrm{Tr}\left(\M{X}\M{\M{\mathcal{L}}}_c\M{X}\Tra\right)$  to classical problem formulations; such regularization incentivizes the recovered signal to be smooth with respect to the underlying data graphs~(\ref{eq:quad}).
The matrices $\M{\mathcal{L}}_r$ and $\M{\mathcal{L}}_c$ denote the graph Laplacians on the rows and columns of $\M{X}$ respectively, and the nonnegative tuning parameters $\gamma_r$ and $\gamma_c$ trade off data fit with smoothness with respect to the row and column geometries encoded in $\M{\mathcal{L}}_r$ and $\M{\mathcal{L}}_c$ respectively. 

\begin{table}[]
\caption{Multi-way Graph regularization formulations}
\footnotesize
\centering
\begin{tabular}{ l c c c }
        & Fidelity term                 & Graph regularizers    & additional constraints  \\ \hline
MCG~\cite{kalofolias2014matrix}     &  $\lVert \mathcal{P}_{\Theta}(\M{Y}-\M{X}) \rVert^2_{\text{F}}$    & $\gamma_r \textrm{Tr}(\M{X}\Tra\M{\mathcal{L}}_r\M{X})+\gamma_c \textrm{Tr}(\M{X}\M{\mathcal{L}}_c\M{X}\Tra)$     &   $\gamma_n\lVert \M{X} \Vert_{\ast} $                        \\ \hline
CFGI~\cite{rao2015collaborative} & $\lVert \mathcal{P}_{\Theta}(\M{Y}-\M{D}\M{X}) \rVert^2_{\text{F}}$ & $ \gamma \left(\textrm{Tr}(\M{D}\Tra\M{\mathcal{L}}_r\M{D})+ \textrm{Tr}(\M{X}\M{\mathcal{L}}_c\M{X}\Tra)\right)$  
 & $ \alpha \lVert \M{D}\rVert_{\text{F}}^2+ \beta \lVert \M{X} \rVert_{\text{F}}^2$ \\ \hline
DGRDL~\cite{yankelevsky2016dual}   & $\lVert \M{Y}-\M{D}\M{X} \Vert^2_{\text{F}} $ & $\gamma_r \textrm{Tr}(\M{D}\Tra\M{\mathcal{L}}_r\M{D})+\gamma_c \textrm{Tr}(\M{X}\M{\mathcal{L}}_c\M{X}\Tra)$ &   $ \quad\quad \lVert \V{x}_i\rVert_0 $                     \\ \hline
T-V Reg~\cite{grassi2017time}   & $\lVert \M{Y}-\M{X} \Vert^2_{\text{F}} $ & $\gamma_r \textrm{Tr}(\M{X}\Tra\M{\mathcal{L}}_G\M{X})+\gamma_c \textrm{Tr}(\M{X}\M{\mathcal{L}}_T\M{X}\Tra)$ &                        \\ \hline
T-V Inpaint~\cite{grassi2017time}   & $\lVert \mathcal{P}_{\Theta}(\M{Y}-\M{X}) \rVert^2_{\text{F}}$ & $\gamma_r \textrm{Tr}(\M{X}\Tra\M{\mathcal{L}}_G\M{X})+\gamma_c \textrm{Tr}(\M{X}\M{\mathcal{L}}_T\M{X}\Tra)$ &                        \\ \hline
Cvx Biclust~\cite{Chi2017a}   & $\lVert \M{Y}-\M{X} \rVert^2_{\text{F}}$ & \begin{tabular}[c]{@{}l@{}} $\gamma_r \sum_{(i,j) \in \mathcal{E}_r} w_{i,j} \lVert \M{X}_{i \cdot} -  \M{X}_{j \cdot} \rVert_2 + $ \\ $\gamma_c \sum_{(i,j) \in \mathcal{E}_c} \tilde{w}_{i,j} \lVert \M{X}_{\cdot i} -  \M{X}_{\cdot j} \rVert_2 $ \end{tabular} &                        \\ \hline
Comani-missing~\cite{mishne2019co}   & $\lVert  \mathcal{P}_{\Theta}(\M{Y}-\M{X}) \rVert^2_{\text{F}}$ & \begin{tabular}[c]{@{}l@{}} $\gamma_r \sum_{(i,j) \in \mathcal{E}_r} \Omega(\lVert \M{X}_{i \cdot} -  \M{X}_{j \cdot} \rVert_2) + $ \\ $\gamma_c \sum_{(i,j) \in \mathcal{E}_c} \Omega(\lVert \M{X}_{\cdot i } -  \M{X}_{\cdot j} \rVert_2) \rVert_2 $ \end{tabular} &                        \\ \hline
FRPCAG~\cite{Shahid2016}     & $\lVert \M{Y}- \M{X} \Vert_{1} $      & $\gamma_r \textrm{Tr}(\M{X}\Tra\M{\mathcal{L}}_r\M{X})+\gamma_c \textrm{Tr}(\M{X}\M{\mathcal{L}}_c\M{X}\Tra)$     &                          \\ \hline
DNMF~\cite{shang2012graph}     & $\lVert \M{Y} - \M{D}\M{X} \Vert^2_{\text{F}} $      & $ \gamma_r \textrm{Tr}(\M{D}\Tra\M{\mathcal{L}}_r\M{D})+ \gamma_c\textrm{Tr}(\M{X}\M{\mathcal{L}}_c\M{X}\Tra)$  &       $ \M{D} \geq 0, \M{X}\geq 0$                    \\ \hline
\\
\vspace{-1cm}
\end{tabular}
\label{table-MWSP}
\end{table}

\begin{figure*}[th]
     \centering
    \includegraphics[width=0.8\linewidth]{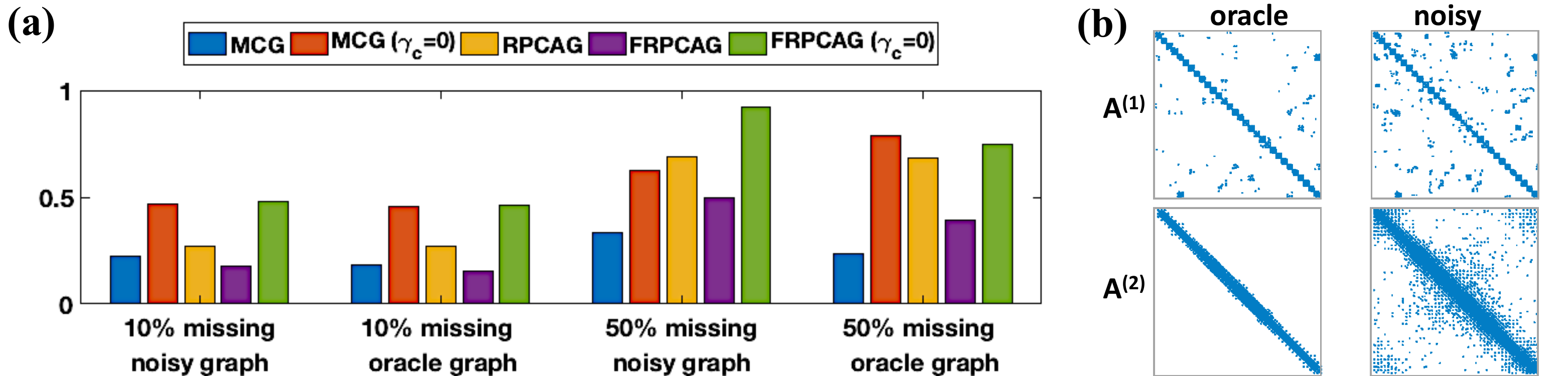}
    \captionof{figure}{\small  Matrix completion on the ORL dataset. (a) Relative error  for $10\%$ and $50\%$ missing values using noisy and oracle graphs. (b) Adjacency matrix of row $\M{A}_r$ and column $\M{A}_c$ graphs for complete data (`oracle') and $50\%$ missing data (`noisy'). }
    \label{fig:mat_complete}
 \end{figure*}

Table~\ref{table-MWSP} presents formulations of these different algorithms; multiple extensions and other methods exist in the literature. 
For the time-vertex framework~\cite{grassi2017time}, the graph on the columns is a temporal graph modeled explicitly with a ring graph Laplacian $\M{\mathcal{L}}_T$.  
The mapping $\mathcal{P}_{\Omega}$ is a projection operator on the set of observed entries $\Theta$ in missing data scenarios.
Methods may differ in their fidelity term minimizing the Frobenius norm for denoising or 1-norm to impart robustness to outliers~\cite{Shahid2016}, and several methods assume a low-rank structure, either with a nuclear norm penalty~\cite{kalofolias2014matrix} or with an explicit low-rank factorization of the data matrix $\M{Y}$ as $\M{D}\M{X}$, sometimes with additional constraints on the factor matrices (non-negativity~\cite{shang2012graph}, sparsity~\cite{yankelevsky2016dual}. A few methods aim to solve a matrix completion problem (see Fig.~\ref{fig:mat_complete}).   
Finally, while most instances of graph regularization rely on the quadratic penalty term 
$\textrm{Tr}\left(\M{X}\Tra\M{\mathcal{L}}_r\M{X}\right) = \sum_{(i,j) \in \mathcal{E}_r} {w}_{i,j}\lVert \M{X}_{i\cdot } -  \M{X}_{j \cdot } \rVert^2_2$, the biclustering formulation in~\cite{Chi2017a,mishne2019co} employs a penalty that is either linear in the $l_2$-norm or concave and continuously differentiable relying on the mapping $\Omega(\lVert \M{X}_{i \cdot} -  \M{X}_{j \cdot} \rVert_2)$.
The motivation there is that convex penalties, either when $\Omega$ is linear or quadratic, do not introduce enough smoothing for small differences and too much smoothing for large differences, resulting in poorer clustering results. 

Typically an alternating optimization algorithm is used to solve the various problems in Table~\ref{table-MWSP}.  
The T-V regularization problem is the only one with a closed form solution given by a joint non-separable low-pass filter (generalizing Tikhonov regularization to the T-V case).
The graph Dual regularized Non-negative Matrix Factorization (DNMF)~\cite{shang2012graph} relies on an alternating optimization scheme for the non-negative factor matrices.
Other solutions are computed with proximal methods such as Alternating-Direction Method of Multipliers (ADMM) to handle multiple regularization terms via variable splitting. 


  Dual-graph regularized approaches have been shown to consistently out-perform their non-regularized or single-graph regularized counterparts across a wide range of applications and domains. 
In Fig.~\ref{fig:mat_complete}(a) we compare several approaches for matrix completion~\cite{kalofolias2014matrix,shahid2015robust,Shahid2016} with single way or multi-way graph regularization on the ORL  dataset with 10\% or 50\% entries missing at random. The ORL~\cite{ORLdata} dataset consists of 300 images of faces (30 people with 10 images per person), which are flattened into 2576 features. We used a row graph that connects similar images together and a column graph that ignores the natural 2D grid geometry and instead considers a wider geometry in the image plane. 
To set $\gamma_r,\gamma_c$, we ran each method for a range of values and selected the result with best performance. 
For comparison to single-way graph regularization, we also set $\gamma_c=0$ in MCG~\cite{kalofolias2014matrix} and FRPCAG~\cite{shahid2015robust} to ignore the graph on the feature (column) space.  
In general, $\gamma_r$ and $\gamma_c$ induce row and column smoothness at different levels and their choice should be driven by the trade-off in the smoothness of the data along the two modes and the aspect ratio of the matrix, or informed by cross-validation.

We report the relative reconstruction error on the missing values, averaged over 10 realizations.
The multi-way graph regularized approaches out-performed their corresponding single-way versions ($\gamma_c=0$) in all cases. Both FRPCAG and MCG always out-performed RPCAG, a single-way graph regularised method. 

\begin{mdframed}[userdefinedwidth=\linewidth,align=center,usetwoside=false,rightmargin=-0.2in,
linecolor=blue,linewidth=1pt,frametitle = {Graph Construction}]
\label{sec:graph}
\small{
A question that arises in graph-based methods is how to construct the graphs themselves.  
In some applications, e.g., social or citation networks, the graph is known a-priori. In transportation or communication networks, vertices represent physical locations (traffic intersections) or sensors (routers in a wifi network), and edges encodes connected locations.
In other settings there is no a-priori graph, and the topology must be learned from the data. We describe common strategies and challenges.

\noindent {\bf Data-driven graph:} One of the most popular ways to construct a graph is from the data itself, for example, using a $k$-nearest neighbor graph with Gaussian kernel weights. For example, in our simulations the row graph weights are $\M{W}^{(1)}_{i,j}=\exp\{-\Vert \M{X}_{i\cdot} - \M{X}_{j\cdot} \Vert^2_2 / \sigma \}$ if rows $i$ and $j$ are $k$-nearest neighbors and zero otherwise, $\sigma$ is bandwidth parameter and we set  $k=7$.
One difficulty that arises is that in the presence of noise, outliers and missing entries, constructing a graph from the data yields a corrupted graph.
 Fig.~\ref{fig:mat_complete}(b) compares a `noisy' graph constructed from the missing data to an `oracle' graph constructed from the original complete data. The noisy graph along the images ($\M{A}^{(1)}$) connects images of different people together while the noisy feature graph ($\M{A}^{(2)}$) loses local pixel geometry. Results in Fig.~\ref{fig:mat_complete}(a) demonstrates that for higher percent of missing values the noisy graph degrades the performance compared to the oracle graph. 

\noindent {\bf Graphs from side information:} Supplementary information can be leveraged to define similarity structure among rows or columns for the purpose of graph construction. 
In some cases, there may be a natural geometry that easily translates into similarity graphs for rows and columns. 
For example, in \cite{loukas2019stationary} the authors constructed the graph among the weather stations using their physical coordinates. In other cases, other supplemental data sets may be leveraged to provide similarity structure among rows or columns. As an example in music recommendation systems, in \cite{Benzi2016} the authors used a publicly available playlist categorization as well as  summary statistics extracted from the audio signal to construct a graph for estimating a latent association matrix between playlists and songs.

\noindent {\bf Graph learning:} In~\cite{yankelevsky2016dual,su2018graph} the graphs on the feature space are learned alongside the signal, by minimizing over $\M{\mathcal{L}}$ in addition to the signal recovery in the optimization problem. For a detailed review see~\cite{mateos2019connecting}.

\noindent {\bf Dynamically varying graphs:} Graphs may not be static, presenting a current challenge in GSP, which is especially acute in time-vertex frameworks which admit time as one of components in the analysis.
Challenges include determining how to identify when a graph needs to be updated, i.e., when the underlying topology has changed. The challenge of accounting for dynamically varying graphs also poses computational questions, namely  what are computationally efficient ways to update graphs within the processing framework that will minimally spawn artifacts at transitions?
} 
\end{mdframed}

\vspace{-0.5cm}
\subsection{Tensor processing on graphs}
\input{tensor_decomp}

\vspace{-0.3cm}
\subsection{Manifold learning on multi-way data}
\label{sec:factoranalysis/manifoldlearning} 
\input{manifold}

%% file: tensor_decomp.tex
\label{sec:factoranalysis/linearfactorization}
\label{sec:tensor_decomp}


A challenge of many well-studied problems in signal processing and machine learning is that algorithm complexity typically grows exponentially when one considers tensors with three or more modes. 
Early multi-way data analysis approaches flattened data tensors to matrices and then applied classical two-way analysis techniques. Flattening, however, obscures higher-order patterns and interactions between the different modes of the data. 
Thus, multilinear tensor decompositions have been the main workhorse in tensor signal processing and data analysis, generalizing the notion of matrix factorizations to higher-order tensors, and have become common in applications such as hyperspectral and biomedical imaging. 

While there is no single generalization of a spectral decomposition for tensors, the two most common tensor decompositions are the CANDECOMP/PARAFAC (CP) decomposition (see \cref{fig:tensor}) 
and the Tucker decomposition \cite{KoldaBader2009}. 
Just as the singular value decomposition can be used to construct a lower-dimensional approximation to a data matrix, finding a coupled pair of lower dimensional subspaces for the rows and columns, these two decompositions can be used to construct lower dimensional approximations to a $D$-way tensor $\T{X} \in \Real^{n_1 \times n_2 \times \cdots \times n_D}$. Under mild conditions, the CP decomposition, which approximates $\T{X}$ by a sum of rank-one tensors, is unique up to scaling and permutations of the columns of its factor matrices \cite{KoldaBader2009}, but the CP factor matrices typically cannot be guaranteed to have orthogonal columns. The Tucker decomposition permits orthonormal factor matrices but in general fail to have unique representations \cite{KoldaBader2009}.
Much of the multi-way literature has focused on improving and developing new tensor factorizations. Graph-based regularizations along modes of the tensor are proving versatile for developing robust tensor and low-rank decompositions~\cite{nie2017graph,zhang2018spatial,shahid2019tensor}, as well as new approaches to problems in higher order data processing such as tensor completion, data imputation, recommendation system, feature selection, anomaly detection, and co-clustering~\cite{li2015multi,sun2018convolutional,ioannidis2018coupled,su2018graph,xie2018graph,Chi2018}---a generalization of biclustering to tensors.
Generalization of these problems to tensors incurs a higher computational cost than the equivalent matrix problems.
Thus multi-way graph-regularized formulations typically combine a low-rank tensor factorization with graph-based regularization along the rows of the factor matrices; for example~\cite{xie2018graph,ioannidis2018coupled} rely on a CP decomposition while~\cite{zhang2018spatial} relies on a Tucker decomposition. 
In~\cite{leonardi2013}, a Tucker decomposition is used within MWGSP, to construct wavelets on multislice graphs in a two-stage approach.

An example of combining tensor decompositions with graph regularization is the following ``low-rank + sparse" model for anomaly detection in internet traffic data~\cite{xie2018graph}:
\begin{equation}
\label{eq:lrs}
\min_{\T{X}, \T{E} \{\M{A}^{(i)}\}_i } \lVert (\T{Y}-\T{E}) - \T{X} \rVert_{\text{F}}^2 + \sum_{i=1}^d \gamma_i \text{Tr} \left(\left(\M{A}^{(i)}\right)\Tra \M{\mathcal{L}}_i   \M{A}^{(i)}\right) \quad 
\text{s.t.} \quad \T{X}=\sum_{i = 1}^R \Vn{a}{1}_i   \circ \Vn{a}{2}_i \circ  \Vn{a}{3}_i, \quad \Vert\T{E} \Vert_0 \leq \epsilon,
\end{equation}
where $\T{Y}$ is a data tensor and 
$\T{E}$ is the tensor of sparse outliers. The equality constraint on $\T{X}$ requires that $\T{X}$ has a rank-$R$ CP decomposition 
where $\Vn{a}{d}_i$ is the $i$th column of the $d$th factor matrix $\Mn{A}{d} \in \Real^{n_d \times R}$ and $\circ$ denotes an outer product. Note that the graph regularization terms in (\ref{eq:lrs}) are applied to the factor matrices $\M{A}^{(i)} \in \mathbb{R}^{n_i \times R}$, reducing the computational complexity of the estimation algorithm. 
Decomposing a data tensor into the sum of a low-rank and sparse components is also used in~\cite{nie2017graph,su2018graph,zhang2018spatial}.

In~\cite{shahid2019tensor}, computational complexity is further reduced by pre-calculating $\M{P}^{(i)}_R$ mode-specific graph Laplacian eigenvectors of rank $R$ from the matricization of the tensor along each mode and using these in solving tensor-robust PCA. The solution relies on projecting the tensor onto a tensor product of the graph basis $\{\M{P}^{(i)}_R\}$, resulting in a formulation to similar to the Tucker decomposition.

Co-clustering assumes the observed tensor is the sum of a ``checkerbox" tensor (under suitable permutations along the modes) and additive noise. 
For example, Chi et al.\@ \cite{Chi2018} propose estimating a ``checkerbox" tensor with the minimizer to a convex criterion. In the case of 3-way tensor, the criterion is
\begin{eqnarray}
\label{eq:coco_objective}
\frac{1}{2}\left\Vert \T{Y} - \T{X} \right\Vert_{\text{F}}^2  +  \gamma
\left[
\sum_{{(i,j) \in \Ec^{(1)}}} w^{(1)}_{ij}  \left\Vert \T{X}_{i::} - \T{X}_{j::} \right\Vert_{\text{F}}  +\;
\sum_{\mathclap{(i,j) \in \Ec^{(2)}}} w^{(2)}_{ij}  \lVert \T{X}_{:i:} - \T{X}_{:j:} \rVert_{\text{F}}  \nonumber 
 + \;
\sum_{\mathclap{(i,j) \in \Ec^{(3)}}} w^{(3)}_{ij}  \lVert \T{X}_{::i} - \T{X}_{::j} \rVert_{\text{F}}
\right],
\end{eqnarray}
where $\Ec^{(d)}$ is a set of edges for the mode-$d$ graph, $\gamma$ is a nonnegative tuning parameter, and $w^{(d)}_{ij}$ is a weight encoding the similarity between the $i$th and $j$th mode-$d$ slices. 
Minimizing the criterion in (\ref{eq:coco_objective}) can be interpreted as 
denoising all modes of the tensor simultaneously via vector-valued graph total-variation. 



%% file: manifold.tex
Tensor factorization can fail to recover meaningful latent variables when nonlinear relationships exist among slices along each of modes. 
Manifold learning overcomes such limitations by estimating \emph{nonlinear} mappings from high-dimensional data to low-dimensional representations (embeddings). 
While GSP uses the eigenvectors of the graph Laplacian as a basis in which to linearly expand graph signals~(\ref{eq:gft}), manifold learning uses the eigenvectors $\V{\psi}_\ell$ themselves as a nonlinear $d$-dimensional map $\Psi$ for the datapoints $\{\mathbf{x}_i \}_i$ as $\Psi : \mathbf{x}_i \rightarrow (\psi_1(i), \ldots, \psi_d(i))$.

A na\"ive strategy to apply manifold learning to the multi-way data is to take the $D$ different matricizations of a $D$-way tensor and construct a graph Laplacian using a generic metric on each of the $D$ modes independently, thereby ignoring the higher-order coupled structure in the tensor. Recent work \cite{Mishne2017,Mishne2016,mishne2019co}, however, incorporate higher-order tensor structure in manifold learning by thoughtfully designing the similarity measures used to construct the mode $k$ graph weights $\M{W}^{(k)}$. 
The co-manifold learning framework can be viewed as blending GSP and manifold learning together and has most recently extended to tensors and the missing data setting~\cite{Mishne2016,mishne2019co}.

From a MWGSP perspective, the key contribution of this line of work is a new metric that is defined between tensor slices as the difference between a graph-based multiscale decomposition of each slice along its  remaining modes; for example the distance between two horizontal slices in a 3-way tensor is 
\begin{equation}
    d\left(\T{X}_{i\cdot\cdot},\T{X}_{j\cdot\cdot}\right) =  \left\Vert \left(\M{M}^{(3)}\Kron \M{M}^{(2)}\right)\vec{(\T{X}_{i::}-\T{X}_{j::})} \right\Vert_1= \left\Vert \vec\left(\M{M}^{(2)}\left(\T{X}_{i::}-\T{X}_{j::}\right) \left(\M{M}^{(3)}\right)\Tra\right) \right\Vert_1,
\end{equation}
where $\M{M}^{(k)}$ is a multiscale  transform in the $k$th mode.
This metric was shown to be a tree-based Earth-mover's distance in the 2D setting~\cite{Leeb2013}. The resulting similarity 
depends on a multi-way multiscale difference between slices, and has been successfully used in practice to construct weighted graphs in multiway data. 
The multiscale decompositions are constructed either from data-adaptive tree transforms~\cite{Mishne2017} or through a series of multi-way graph-based co-clustering solutions~\cite{mishne2019co}.